\documentclass{article}
\usepackage{PRIMEarxiv}
\usepackage[utf8]{inputenc} 
\usepackage[T1]{fontenc}    
\usepackage{hyperref}       
\usepackage{url}            
\usepackage{booktabs}       
\usepackage{amsfonts}       
\usepackage{nicefrac}       
\usepackage{microtype}      
\usepackage{fancyhdr}       
\usepackage{graphicx}      
\graphicspath{{media/}}     

\title{CipherMind:The Longest Codebook in the World}

\author{
  Ming Nie*, Zhixiong Yang*,Bingsheng Wei \\
  Fudan University \\
  ShangHai,China\\
}
\begin{document}
\maketitle

\begin{abstract}
 In recent years, the widespread application of large language models has inspired us to consider using inference for communication encryption. We therefore propose CipherMind, which utilizes intermediate results from deterministic fine-tuning of large model inferences as transmission content. The semantic parameters of large models exhibit characteristics like opaque underlying implementations and weak interpretability, thus enabling their use as an encryption method for data transmission. This communication paradigm can be applied in scenarios like intra-gateway transmission, and theoretically, it can be implemented using any large model as its foundation.
\end{abstract}


\section{Introduction}
The encrypted transmission of information has become very common in the Internet era. The wide application of large language models\cite{1zhao2023survey,2vaswani2017attention} has led us to consider the possibility of using large models for encryption in the field of cryptography. Due to the increasing number of parameters in large models year by year, the gradual complexity of model functions, and the lack of research on parameter semantics\cite{1zhao2023survey}, the problem of interpretability of intermediate results in large models is becoming increasingly difficult to solve. Based on this premise, we believe that for two different inferences of the same large model with different contexts, if we randomly extract the intermediate results of any layer as ciphertext each time, these two sets of ciphertexts are indistinguishable. Therefore, we can attempt to use the output of the intermediate layer of the fine-tuned model as the transmitted information to achieve encrypted information transmission.

We propose CipherMind, a means of encrypting information based on the black-box process of macromodeling. To achieve this goal, we use Qwen2.5-0.5B-Instruct to perform a “heterogeneous twin” fine-tuning method. That is, the same dataset and random seeds are used for fine-tuning to ensure that the models obtained after fine-tuning are the same and the inference is the same.

\begin{figure} 
    \centering
    \includegraphics[width = 10cm,height = 12.5cm]{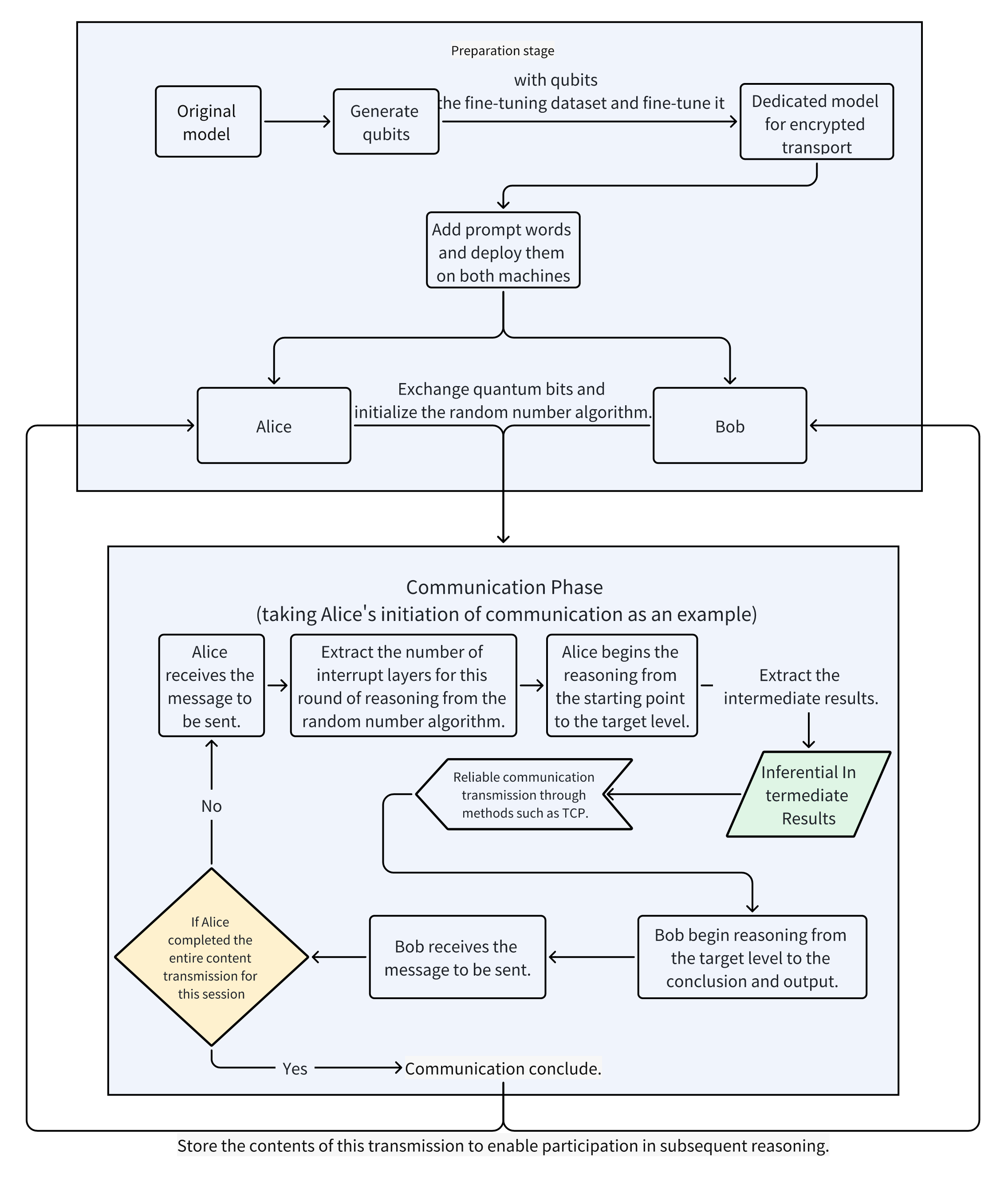}
    \caption{CipherMind Architecture}
\end{figure}

The contributions of this work are as follows:
\begin{itemize}
    \item Implementation of \textbf{deterministic fine-tuning method} that ensures identical post-tuning parameters across different machines, achieving \textbf{"Distant Twinness"}
    \item CipherMind encrypted communication model is proposed to encrypt and decrypt information by means of middle layer output reception, and combined with CBC encryption process to ensure security
\end{itemize}
\section{Background and Motivation}
\subsection{Background}

Encryption, is the method of transforming information to be transmitted (plaintext) into another unreadable form of information (ciphertext). While encryption does not prevent information from being stolen or intercepted, it does make the content invisible to an attacker. Ideally, only the authorized party can convert the ciphertext into plaintext (decryption). In the scenario of encrypted message transmission in a network, encryption is generally accomplished through pseudo-random numbers, encryption algorithms, etc. Therefore, theoretically, the attacker can decrypt the key by brute-force decryption, and an ideal encryption method does not exist. The actual encryption algorithm only needs to satisfy the computational security \cite{3katz2007introduction}.

\subsection{Motivation}
With the rise of AI technology, arithmetic power has been dramatically increased \cite{1zhao2023survey}, which has made algorithms that used to be secured by virtue of high computational power vulnerable to attack. Combined with the fact that large language models require a large amount of arithmetic power for both training and use, we pondered whether this feature could be exploited to construct an encryption method whose computational complexity matches current high-computing-power techniques, making it impossible for attackers to crack even if they consume a large amount of resources.

One of the most important features of the large language model is the non-interpretability of its parameters and intermediate outputs\cite{4rane2023chatgpt}. We believe that such non-interpretable intermediate outputs can be used as a kind of cipher text for transmission of information, and combined with the reasoning process of the large language model, the information to be transmitted will be inputted in the form of prompt, and the intermediate layer will be intercepted as the encryption result in a random manner and passed to the receiver to complete the encryption process. The intermediate layer is “randomly” intercepted as the result of encryption and passed to the receiver for parsing to complete the encryption process. There is no effective way to decode, mimic, or even modify this method, so it may have a wide range of application prospects.

Thus, we had the following motivation for building CipherMind:

\begin{itemize}
    \item In the course of conducting research and learning, we found that many cases of deep integration with AI technologies are emerging in the field of data encryption. However, no similar work has yet emerged that integrates with AI reasoning intermediate processes for encryption.
    
    \item As arithmetic power continues to expand, new high-computer-power encryption methods are needed; traditional algorithms have advantages such as rigor and self-consistency, but they still have weaknesses in terms of interpretability, and traditional algorithms have the potential to become obsolete.
    
    \item The intermediate results of large model inference have the advantages of being weakly interpretable, randomized, and not easily imitated or tampered with, and thus have the possibility of being deeply combined with cryptographic algorithms.
\end{itemize}

\section{Design}

In this section, we introduce the design of CipherMind, followed by several key technical components including (1) fine-tuning the design (2) heterogeneous twinning (3) CBC encryption. While presenting these key parts, we will present relevant threat models as a way to clarify the need for these techniques.

\subsection{CipherMind Overview}
    The structural framework and specific operation of CipherMind are shown below.

\begin{figure}
    \centering
    \includegraphics[width = 10cm,height = 8cm]{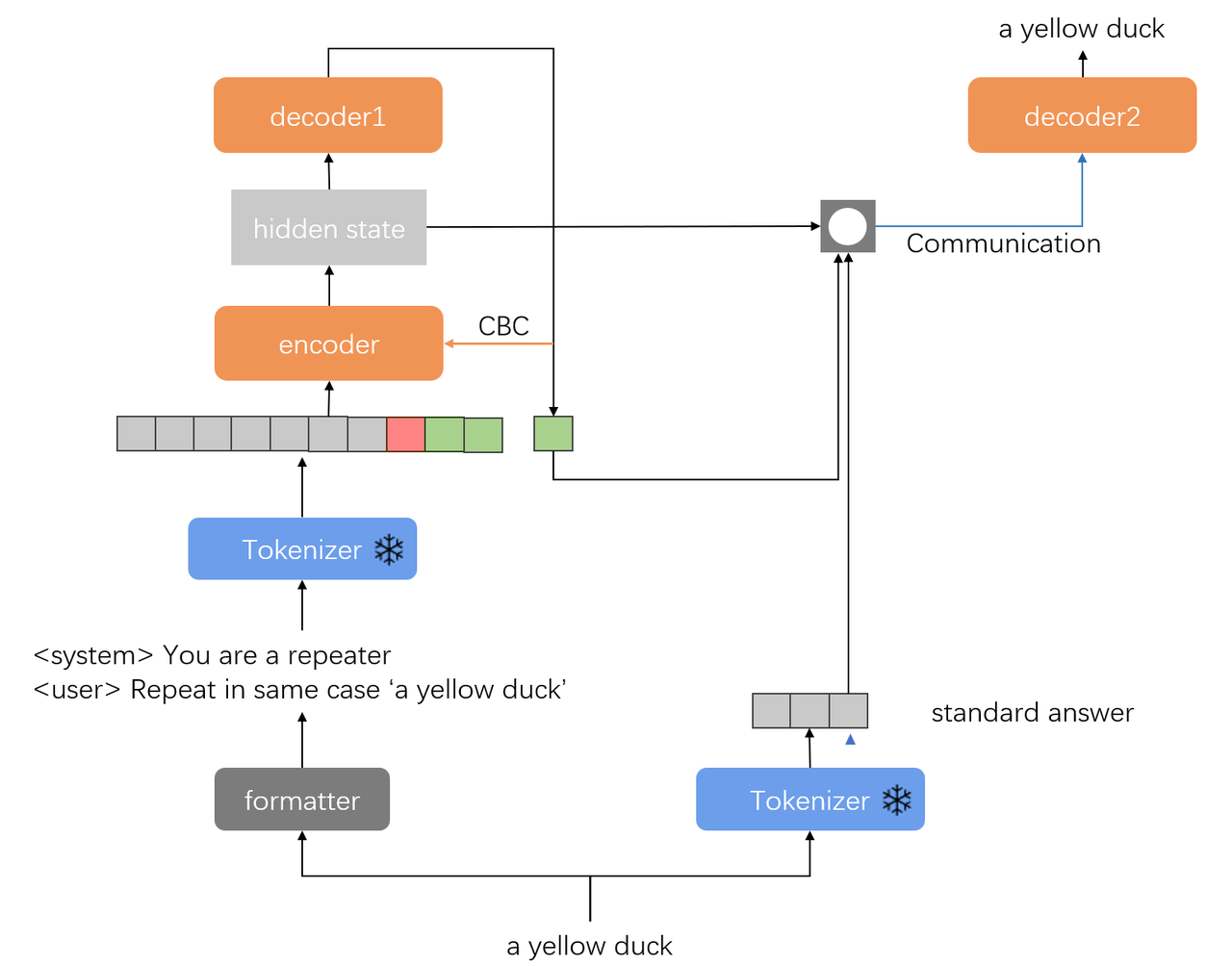}
    \caption{For the input text, it is first added to a predefined template that requires the model to repeat the specified content, after which it is generated token by token. For each token generated, the desired content is compared and if it matches, the output of the middle layer is transmitted until the next token is <eos>. At the same time, each output token is updated with a CBC encryption for the next middle layer selection.}
\end{figure}

CipherMind Workflow
CipherMind's design philosophy can be summarized in the following three points:
\begin{enumerate}
    \item Fine-tuning design
    CipherMind will be designed to be deterministically fine-tuned using a small parameter model, making it difficult for an attacker to get meaningful results by putting the middle layer output of the fine-tuned model into the original model even if the attacker knows what model the encryption is using (see Section 5.3). Specifically, one can construct a set consisting of 128 small data sets and pick a subset with random numbers before the communication begins; ultimately the attacker must brute-force exhaustively enumerate $2^{128}$ combinations to get the correct fine-tuned version.
   
   \item Heterogeneous twinning
   
    In order to avoid the problem of parameter interception after transmission fine-tuning, the method of synchronized heterogeneous fine-tuning is used here. By ensuring that the same dataset is used and the same random seed is used, the final parameters fine-tuned by the two communicating parties are identical. In addition, the random seed setting in the inference process ensures the complete consistency of the intermediate process, so as to ensure that the intermediate results of the sender can get the same results as expected at the receiver, thus realizing the “heterogeneous twin”.

   \item CBC Encryption
   
    In addition to the inherent non-interpretability of the output of the model intermediate layer, the “randomness” of the number of intermediate layers is introduced in the communication process. To be precise, the selection of intermediate layers is encrypted using a token-based CBC cipher. For the generation of a sentence, every time a token is output, the number of layers in the next output can be easily changed according to all the previously output tokens, and the “heterogeneous twinning” ensures that the receiver and the sender can be adjusted in the same way, which further guarantees the security of the encryption algorithm.

\end{enumerate}

\section{Experiment}

In this section, we present our simulation experiments and results based on our proposed theory and architecture.

\subsection{Experimental settings}

    All the base models used in this section are Qwen2.5-0.5B-Instruct models, which were all fine-tuned using the lora fine-tuning method. When performing the fine-tuning, we found that the number of fine-tuning steps during the fine-tuning process has a large impact on the final performance of the model, so we used different numbers of fine-tuning steps to fine-tune the base models to different extents and experimented with them in order to demonstrate the effectiveness and stability of the CipherMind design. The datasets used for fine-tuning were all customized and modified squad\_v2 datasets (with cue-word-output pairs based on our task injected into them at a 10:1 ratio in order to achieve our “repetition” task). Details of the code can be found in our github repository.

\subsection{Correct Transmission Capability Experiment}

    The following experiments were conducted to investigate the optimal fine-tuning parameters for the model and to test the transmission capability of the model. Random strings were used in all the experiments, and if the final output contained the target string and successfully output <eos>, the transmission was considered successful.

    \begin{figure}
    \centering
        \includegraphics[width = 14cm,height = 7cm]{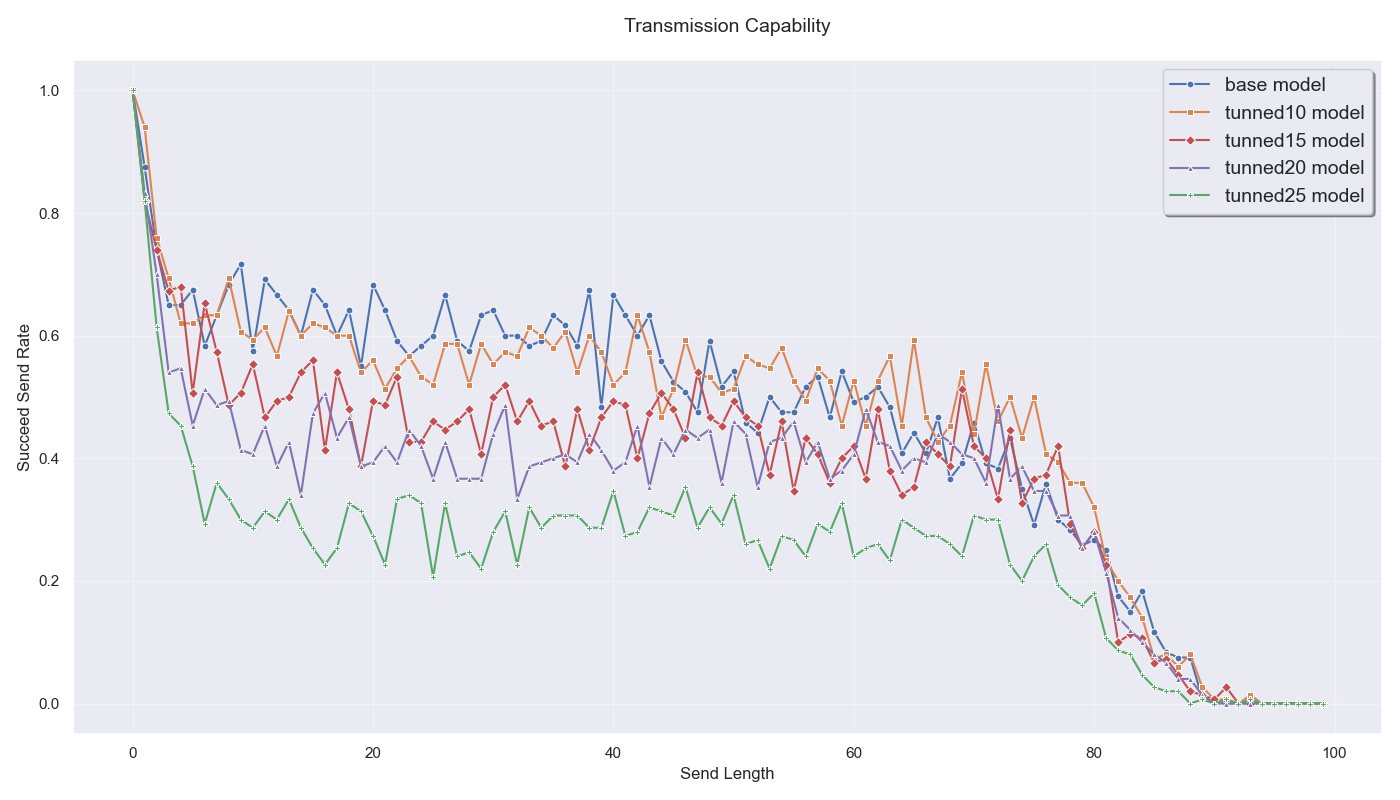}
        \caption{Information Transfer Capability of Models with Different Levels of Fine-Tuning}
    \end{figure}
   The experimental results show that when the fine-tuning parameter is set to 20 or less, all the models have almost no loss compared to the original model, and even have stronger transmission ability for longer length content. However, as the transmission length increases, correctness transmission becomes more and more difficult. We believe that this may be due to:
    \begin{itemize}
        
        \item The model itself has small parameters, which creates a limitation on the model's capabilities;
        
        \item The quality of the dataset is average and there is some room for optimization of the fine-tuning instruments;
    \end{itemize}

\subsection{crash experiment}
    This experiment is intended to simulate the stability of the architecture under attack by a CPA-capable attacker. In order not to let the semantic information influence the model, random strings are still transmitted using the same implementation as the correct transmission. both the sender and the attacker use the CipherMind implementation, but the information obtained by the attacker is different each time; and the sender uses the fine-tuned model, while the attacker uses the original model.

    \begin{figure}
    \centering
        \includegraphics[width = 14cm,height = 7cm]{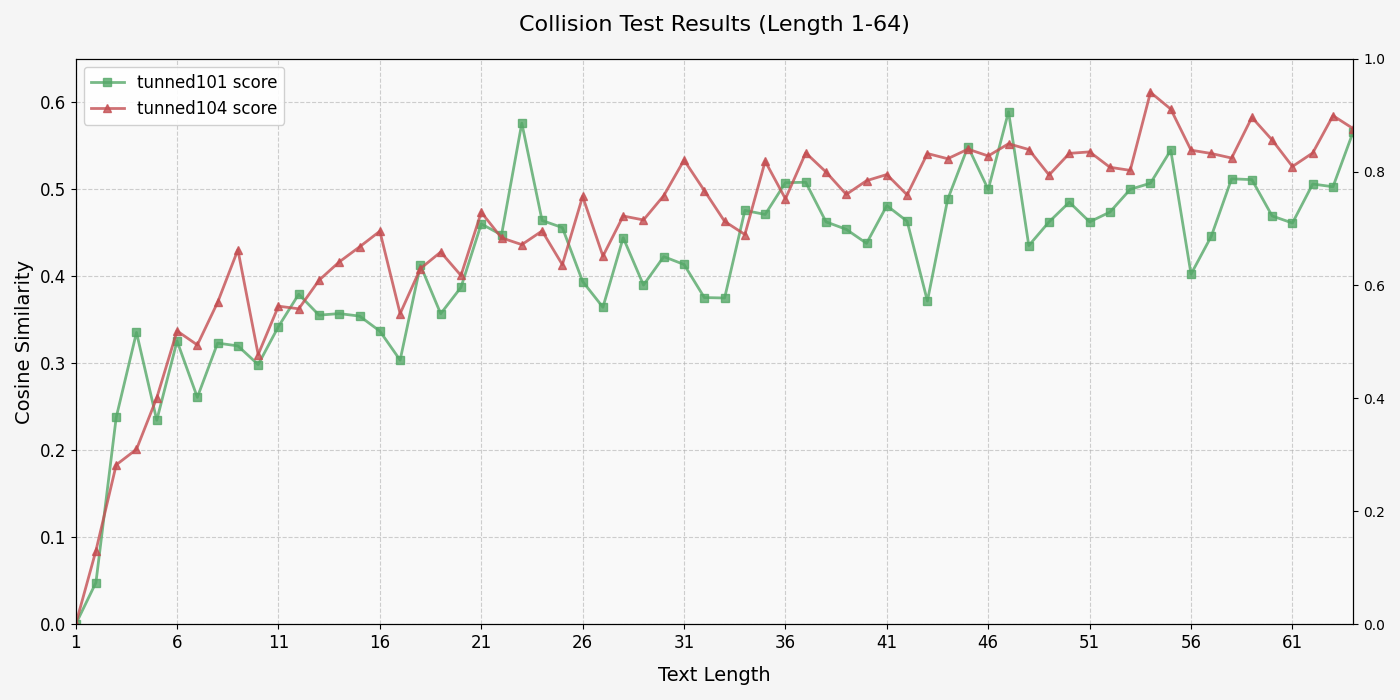}
        \caption{Comparison of collision probabilities after ablation of random layers}
    \end{figure}
    
    Taking the results of the correctness experiments as a reference, when conducting the collision experiments, we only choose strings with a length of 64 characters and below for the experiments. First, we chose to ablate the effect of the number of random layers, attacker next can not know sender's pseudo-random number generator with the current seed; in the figure above, it can be seen that if this is ablated, then the collision rate will rise by about 20\%. We speculate that this may be due to:
    \begin{itemize}
        \item The fine-tuning brings about a change in the model's intermediate parameters, which by itself eliminates part of the collision probability;
        \item The Qwen2.5-0.5B-Instruct model goes through 27 intermediate layers per inference, and the collision probability of random layer elimination should be related to the model layer size.
    \end{itemize}
    
    \begin{figure}
    \centering
        \includegraphics[width = 14cm,height = 7cm]{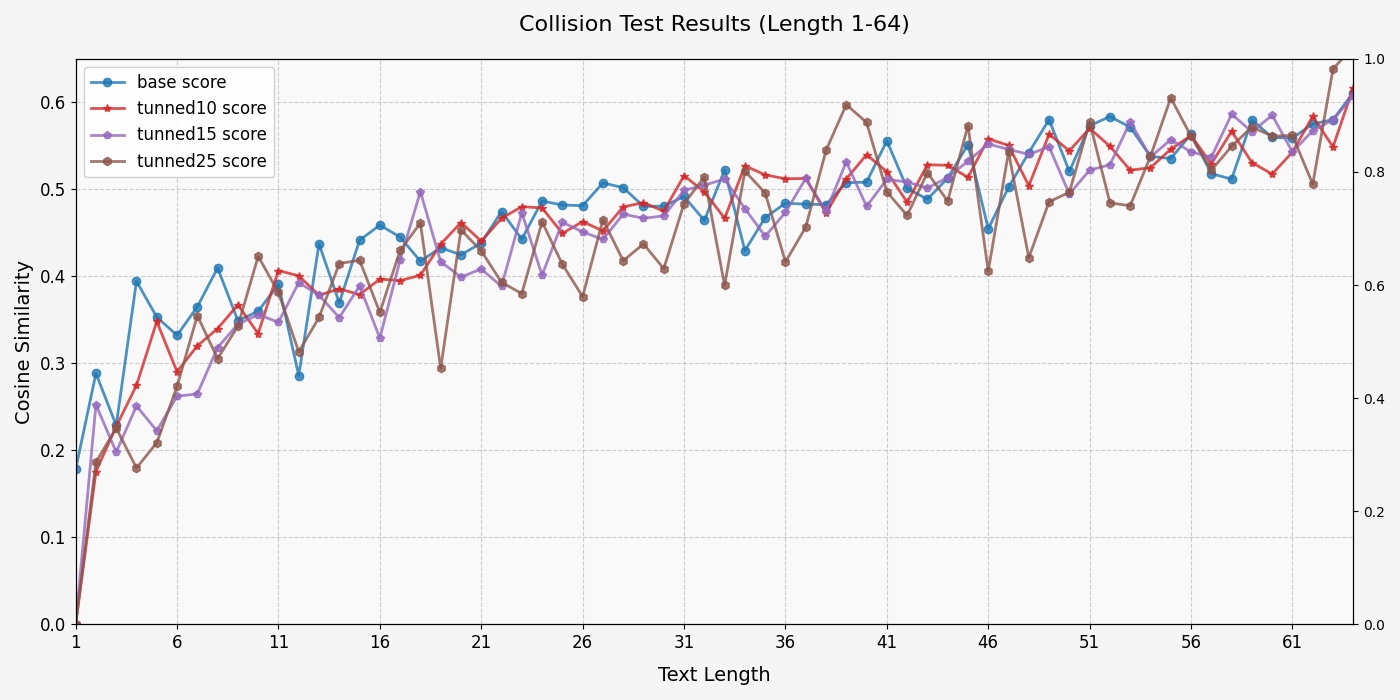}
        \caption{Comparison of collision probability after ablation fine-tuning}
    \end{figure}
    
    Second, we compared the effect of different intensities of fine-tuning on the collision rate. In the above figure, BASE SCORE indicates that the original model was used as the SENDER, and TUNNED10 and TUNNED20 indicate different degrees of fine-tuning the model. Compared to the original model, the models with greater degrees of fine-tuning have lower cosine similarity when performing collision experiments; however, the decrease in collision rate is not significant, and we speculate that it might be due to:
    \begin{itemize}
        \item The number of model parameters is small, and the model can afford to be fine-tuned to cause lower impacts.
        \item Due to the attentional mechanism, the model increases the failure rate when it receives content containing a question or an imperative tone.
        \item In this experiment, random strings were used for transmission in order to try to eliminate the effect of the semantic level, and the final cosine similarity calculations were performed on a character-by-character basis; with the addition of semantics, the effect of fine-tuning on the results should be even more significant.
    \end{itemize}

Overall, it is worthwhile to make a trade-off between the size of the model parameters and the fine-tuning parameters according to the actual usage scenarios, so as to realize the greater potential of CipherMind.

\section{Open Exploration}
\subsection{Practical application concepts}
    According to the P2P encryption features of CipherMind, one can choose to implement a cryptographic network inside the gateway:after all members agree to the protocol, the pre-fine-tuning macromodel is deployed uniformly; whenever a new line is to be established, random numbers are exchanged through methods such as RSA to generate a set of heterogeneous twin macromodels locally on both sides. A similar mechanism can be used to liaise between different gateways.

\subsection{Ideal encryption strength}
    There are two main encryption methods in the CipherMind encryption process:
    (1)Fine-tuning encryption before deployment
    (2)Dynamic encryption during communication
\begin{enumerate}
    \item Fine-tuning encryption before deployment
    
        When communication is first established, a series of keys are exchanged between the two parties via RSA and are used to trade off among 128 fine-tuned datasets to create a model space of size $2^{128}$. Ideally, these models do not interfere or collide with each other. Assuming the eavesdropper attempts a brute-force crack, each field is a key that has to be fine-tuned once, thus imposing unrealistic arithmetic requirements.
    
    \item Dynamic encryption during communication
    
       In order to add another insurance policy on top of fine-tuning, CipherMind considers dynamic encryption after the communication is established by utilizing a vectorized storage of the communication history in the form of Retrieval Augmented Generation (RAG)\cite{5lewis2020retrieval} and so on. Under this premise, an eavesdropper who succeeds in cracking the key will also need access to all the communication history for a successful attack, which also allows CipherMind to develop forward secrecy.

\end{enumerate}

     Based on the above two approaches, we believe that this encryption can fulfill the computational security requirements.

\subsection{Potential issues and challenges}

\begin{enumerate}
    \item Deployment cost issues
    The memory footprint of the smallest open-source models currently on the market is still measured in gigabytes (GB); while the lora\cite{6hu2022lora} fine-tuning method supports storing parameters individually, fine-grained deployment between users is still a significant burden on memory and arithmetic. These deployment costs pose a challenge to the widespread deployment of CipherMind.
    \item Fine-tuning dataset selection issues
    Currently there is a lack of research on quantitative computation of semantic similarity between datasets, and the degree of interference of such similarity on the results after fine-tuning also lacks quantitative data support; according to our test, the performance will drop dramatically when the model suffers from catastrophic forgetting, so selecting high-quality datasets to mitigate this phenomenon is also an important issue. How to construct a generalized computational method has selected the optimal collection of datasets can be the focus of subsequent research.
\end{enumerate}

\section{conclusions}

    In this paper, we propose CipherMind, a cryptographic communication framework based on large model reasoning, and conduct preliminary practice based on the ideas therein. After our explorations and experiments, this framework has high cryptographic potential and application prospects, and its implementation is not difficult, the cost is easy to be accepted, and the ideal cryptographic strength is very exciting, especially in distributed computing and encryption scenarios may have very strong performance.

\bibliographystyle{unsrt}
\bibliography{references}

Github repository URL:\url{https://github.com/yourusername/CipherMind}
\end{document}